\documentclass[onecolumn,12pt,journal,draftclsnofoot,a4paper,oneside]{IEEEtran}
\usepackage{bbm}
\usepackage{amsfonts}
\usepackage{mathrsfs}
\usepackage{multicol}
\usepackage{cite}
\usepackage{graphicx}
\usepackage{amsmath}
\usepackage{amssymb}
\usepackage{algorithm}
\usepackage{algorithmic}

\ifCLASSINFOpdf

\else

\fi

\begin{document}

\title{\large{Efficiency Resource Allocation for Device-to-Device
Underlay Communication Systems: A Reverse Iterative Combinatorial
Auction Based Approach }\thanks{This paper is partially supported by
US NSF CNS-1117560, ECCS-1028782, CNS-0953377, CNS-0905556, and
Qatar National Research Fund.}}

\author{\IEEEauthorblockN{\normalsize{Chen Xu}\IEEEauthorrefmark{1}, \normalsize{Lingyang Song}\IEEEauthorrefmark{1},
\normalsize{Zhu Han}\IEEEauthorrefmark{2}, \normalsize{Qun
Zhao}\IEEEauthorrefmark{3}, \normalsize{Xiaoli
Wang}\IEEEauthorrefmark{3}, \\ \normalsize{Xiang~Cheng}\IEEEauthorrefmark{1}, and \normalsize{Bingli~Jiao}\IEEEauthorrefmark{1}} \\
\IEEEauthorblockA{\IEEEauthorrefmark{1}\normalsize{Peking University, Beijing, China.}\\
\IEEEauthorrefmark{2}\normalsize{University of Houston, Houston, USA.}\\
\IEEEauthorrefmark{3}\normalsize{DoCoMo Beijing Communications
Laboratories, China.}}}

\maketitle
\begin{abstract}
Peer-to-peer communication has been recently considered as a popular
issue for local area services. An innovative resource allocation
scheme is proposed to improve the performance of mobile
peer-to-peer, i.e., device-to-device (D2D), communications as an
underlay in the downlink (DL) cellular networks. To optimize the
system sum rate over the resource sharing of both D2D and cellular
modes, we introduce a reverse iterative combinatorial auction as the
allocation mechanism. In the auction, all the spectrum resources are
considered as a set of resource units, which as bidders compete to
obtain business while the packages of the D2D pairs are auctioned
off as goods in each auction round. We first formulate the valuation
of each resource unit, as a basis of the proposed auction. And then
a detailed non-monotonic descending price auction algorithm is
explained depending on the utility function that accounts for the
channel gain from D2D and the costs for the system. Further, we
prove that the proposed auction-based scheme is cheat-proof, and
converges in a finite number of iteration rounds. We explain
non-monotonicity in the price update process and show lower
complexity compared to a traditional combinatorial allocation. The
simulation results demonstrate that the algorithm efficiently leads
to a good performance on the system sum rate.
\end{abstract}
\newpage
\section{Introduction}
As one of next-generation wireless communication systems, Third
Generation Partnership Project (3GPP) Long Term Evolution (LTE) is
committed to provide technologies for high data rates and system
capacity. Further, LTE-Advanced (LTE-A) was defined to support new
components for LTE to meet higher communication demands
\cite{Ref:1}. Local area services are considered as popular issues
to be improved, and by reusing spectrum resources local data rates
have been increased dramatically. However, the unlicensed spectrum
reuse may bring inconvenience for local service providers to
guarantee a stable controlled environment, e.g., ad hoc network
\cite{Ref:adhoc}, which is not in the control of the base station
(BS) or other central nodes.
Hence, accessing to the licensed spectrum has attracted
much attention.

Device-to-Device (D2D) communication is a technology component for
LTE-A. The existing researches allow D2D as an underlay to the
cellular network to increase the spectral efficiency
\cite{Ref:1,Ref:2}. In D2D communication, user equipments (UEs)
transmit data signals to each other over a direct link using the
cellular resources instead of through the BS, which differs from
femtocell \cite{Ref:femtocell} where users communicate with the help
of small low-power cellular base stations. D2D users communicate
directly while remaining controlled under the BS. Therefore, the
potential of improving spectral utilization has promoted much work
in recent years \cite{Ref:3,Ref:4,Ref:5,Ref:6,Ref:7,Ref:8}, which
shows that D2D can improve system performances by reusing cellular
resources. As a result, D2D is expected to be a key feature
supported by next generation cellular networks.

Although D2D communication brings improvement in spectral efficiency
and makes large benefits on system capacity, it also causes
interference to the cellular network as a result of spectrum
sharing. Thus, an efficient interference coordination must be
formulated to guarantee a target performance level of the cellular
communication. There exists several work about the power control of
D2D UEs for restricting co-channel interference
\cite{Ref:1,Ref:2,Ref:9,Ref:10}. The authors in \cite{Ref:11}
utilized MIMO transmission schemes to avoid interference from
cellular downlink to D2D receivers sharing the same resources, which
aims at guaranteeing D2D performances. Interference management both
from cellular to D2D communication and from D2D to cellular networks
are considered in \cite{Ref:12}. In order to further improve the
gain from intra-cell spectrum reuse, properly pairing the cellular
and D2D users for sharing the same resources has been studied
\cite{Ref:13,Ref:14}. The authors in \cite{Ref:14} proposed an
alternative greedy heuristic algorithm to lessen interference to the
primary cellular networks using channel state information (CSI). The
scheme is easy-operated but cannot prevent signaling overhead. In
\cite{Ref:15}, the resource allocation scheme avoids the harmful
interference by tracking the near-far interference, identifies the
interfering cellular users, and makes the uplink (UL) frequency
bands efficiently used. Also, the target is to prevent interference
from cellular to D2D communication. In \cite{Ref:16}, the authors
provided analysis on optimum resource allocation and power control
between the cellular and D2D connections that share the same
resources for different resource sharing modes, and evaluated the
performance of the D2D underlay system in both a single cell
scenario and the Manhattan grid environment. Then, the schemes are
to further optimize the resource usage among users sharing the same
resources. Based on the aforementioned work, it indicates that by
proper resource management, D2D communication can effectively
improve the system throughput with the interference between cellular
networks and D2D transmissions being restricted. However, the
problem of allocating cellular resources to D2D transmissions is of
great complexity. Our works differ from all mentioned above in that
we consider a scheme to maximize the system sum rate by allowing
multiple pairs share one cellular user's spectrum resource.

Game theory offers a set of mathematical tools to study the complex
interactions among interdependent rational players and to predict
their choices of strategies \cite{Ref:17}. In the present
researches, game theory including a large number of different game
methods are used to analyze resource allocation problems, such as
power and wireless spectrum allocations in communication networks
\cite{Ref:18}, resource management in grids \cite{Ref:19}, and
distributed resource coordination in mega-scale container terminal
\cite{Ref:20}. In \cite{Ref:18}, the authors proposed a sequential
auction for sharing the wireless resource, which is managed by a
spectrum broker that collects bids and allocates discrete resource
units using a sequential second-price auction. A combinatorial
auction model for resource management was introduced in
\cite{Ref:19,Ref:20}. The combinatorial auction-based resource
allocation mechanism allows an agent (bidder) to place bids on
combinations of resources, called ``packages", rather than just
individual resource unit.

Actually, the combinatorial auctions (CAs) have been employed in a
variety of industries for, e.g., truckload transportation, airport
arrival and departure slots, as well as wireless communication
services. The benchmark environment of auction theory is the private
value model, introduced by Vickrey (1961), in which one bidder has a
value for each package of items and the value is not related to the
private information of other bidders \cite{Ref:21}. Much of work has not recognized that bidders care in complex
ways about the items they compete. The CAs motivate bidders to fully
express their preferences, which is an advantage in improving system
efficiency and auction revenues. Up to that point, our interest is
to apply the CA game in solving arbitrary D2D links reusing the same
cellular frequency bands with the purpose of optimizing the system
capacity.

However, it exists a series of problems and challenges in CAs, such
as pricing and bidding rules, the winner determination problem (WDP)
which, as mentioned in the literature, leads to the NP-hard
allocation problem. Therefore, we focus on the evolution mechanisms
named iterative combinatorial auctions (I-CAs) \cite{Ref:22,Ref:23}.
In I-CAs, the bidders submit multiple bids iteratively, and the
auctioneer computes provisional allocations and ask prices in each
auction round.

In this paper, we study an effective spectrum resource allocation
for D2D communication as an underlay to further improve system
efficiency based on the I-CA. The whole system consists of the BS,
multiple cellular users that receive signals from the BS, and
multiple D2D pairs that communicate with respective receivers using
licensed spectrum resources. Considering that interference
minimization is a key point and multiple D2D pairs sharing the same
resources can bring large benefits on system capacity, we formulate
the problem as a reverse I-CA game. That means, the resources as the
bidders compete to obtain business, while D2D links as the goods or
services wait to be sold. By this way, the packages of D2D pairs are
auctioned off in each auction round. Furthermore, we investigate
some important properties of the proposed resource allocation
mechanism such as cheat-proof, convergence and price-monotonicity.
Part of our work has been published in \cite{Ref:ICCpaper}, which
introduces a sequential second price auction as the allocation
mechanism for D2D underlay communication, and explains the detailed
algorithm using an N-ary tree. In this work, we further reduce the
computational complexity and apply our scheme to WINNER II channel
models \cite{Ref:WINNER} which contain a well-known indoor scenario.
The simulation results show that the auction algorithm leads to a
good performance on the system sum rate, and provides high system
efficiency while has lower complexity than the exhaustive search
allocation.

The rest of the paper is organized as follows: In Section
\ref{sec:system_model}, we describe the system model of the D2D
communication underlaying cellular network, and give the explanation
and expression of the system sum rate. The primary problem is
formulated in Section \ref{sec:prob_form}. In Section
\ref{sec:algorithm}, the resource allocation algorithm based on a
reverse I-CA is proposed. In Section \ref{sec:analysis}, the main
properties of the proposed algorithm are investigated. In Section
\ref{sec:simulations}, we present the numerical simulation results
and relevant analysis on the system sum rate, algorithm efficiency,
and properties. Finally, we draw the conclusions in Section
\ref{sec:conclusions}.

\section{System Model}\label{sec:system_model}
In this section, we introduce the system model for D2D underlay
communication. The scenario of multiple D2D and cellular users is
first described, and then, the expression of system sum rate is
given.

\subsection{Scenario Description}

\begin{figure}[h!]
\centering
\includegraphics[height=3.7in]{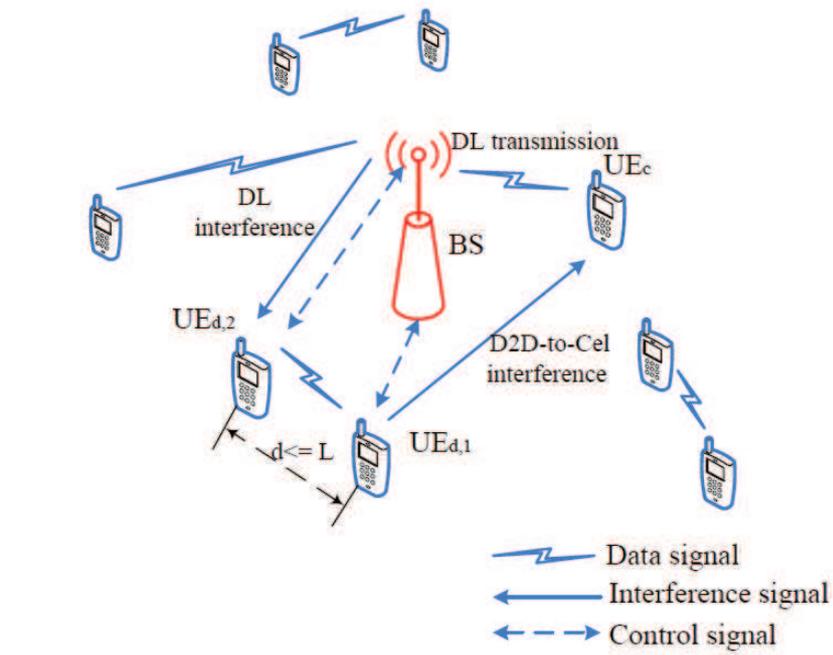}
\caption{System model of D2D communication underlaying cellular
networks with downlink resource sharing.} \label{fig:1}
\end{figure}

A model of a single cell with multiple users is considered. As shown
in Fig. \ref{fig:1}, UEs with data signals between each other are in
the D2D communication mode while UEs that transmit data signals with
the BS keep in the traditional cellular mode. Each user is equipped
with a single omnidirectional antenna. The locations of cellular
users and D2D pairs are randomly set and traversing the whole cell.
Without loss of generality, we employ the uniform distribution to
describe the user locations which is proposed for system simulation
in \cite{Ref:26}. Notice that from stochastic geometry with for
Poisson distributions, the users are uniformly located as well if
the number of users is known \cite{Ref:Jeff}. For simplicity and
clarity, we illustrate co-channel interference scenario involving
three UEs (UE$_c$, UE$_{d,1}$ and UE$_{d,2}$), and omit the
interference and control signal signs among others. UE$_c$ is a
traditional cellular user that is distributed uniformly in the cell.
UE$_{d,1}$ and UE$_{d,2}$ are close enough to satisfy the distance
constraints of D2D communication, and at the same time they also
have communicating demands. One member of the D2D pair UE$_{d,1}$ is
distributed uniformly in the cell, and the position of the other
member UE$_{d,2}$ follows a uniform distribution inside a region at
most $L$ from UE$_{d,1}$.

The existing researches \cite{Ref:15,Ref:16} confirm that with power
control or resource scheduling mechanism, the inter-cell
interference can be managed efficiently. Therefore, we place an
emphasis on the intra-cell interference, which is due to resource
sharing of D2D and cellular communication. Generally speaking, the
session setup of D2D communication requires the following steps
\cite{Ref:1}:
\begin{enumerate}
\item A request of communicating is initiated by one UE pair.
\item The system detects traffic originating from and destined to the
UE in the same subnet.
\item If the traffic fulfills a certain
criterion (e.g., data rate), the system considers the traffic as the
potential D2D traffic.
\item The BS checks if D2D communication
offers higher throughput.
\item If both UEs are D2D capable and D2D
communication offers higher throughput, the BS may set up a D2D
bearer.
\end{enumerate}
The cross-layer processes of resource control can be contained in
the above steps, and be generally summarized as: the transmitters
(both cellular and D2D users) send detecting signals. Then CSI would
be obtained by corresponding receivers and be feedback to the
control center (e.g. the BS). The power control and spectrum
allocation are conducted based on certain principles. Finally, the
BS sends control signals to users according to allocation results.

Even if the D2D connection setup is successful, the BS still
maintain detecting if UE should be back to the cellular
communication mode. Furthermore, the BS maintains the radio resource
control for both cellular and D2D communication. Based on these
communication features, our work mainly focuses on assigning
cellular resources to D2D communication.

In this paper, we consider a scenario of sharing downlink (DL)
resource of the cellular network as shown in Fig.~\ref{fig:1}. We
assume UE$_{d,1}$ is the transmitter of the D2D pair sharing the
same sub-channel with the BS, and thus, UE$_{d,2}$ as the D2D
receiver receives interference from the BS. Also, the cellular
receiver UE$_c$ is exposed to interference from UE$_{d,1}$. In
addition, the D2D users feed back the CSI to the BS, whereas the BS
transmits control signals to the D2D pair, in the way that the
system achieves D2D power control and resource allocation.

During the DL period of the cellular system, both cellular and D2D
users receive interference as they share the same sub-channels.
Here, we assume that any cellular user's resource blocks (RBs) can
be shared with multiple D2D pairs and each pair can use more than
one user's RBs for transmitting. We assume the numbers of cellular
users and D2D pairs in the model are $C$ and $D$, respectively.
During the DL period, the BS transmits signal $x_c$ to the $c$-th
($c=1,2,...,C$) cellular user, and the $d$-th ($d=1,2,...,D$) D2D
pair uses the same spectrum resources transmitting signal $x_d$. The
received signals at UE $c$ and D2D receiver $d$ are written as
\begin{equation}
y_c  = \sqrt {P_B } h_{Bc} x_c  + \sum\limits_d {\beta _{cd} \sqrt
{P_d } h_{dc} x_d }  + n_c,
\end{equation}
\begin{equation}
y_d  = \sqrt {P_d } h_{dd} x_d  +  \sqrt {P_B } h_{Bd} x_c
 +  \sum\limits_{d'} {\beta _{dd'}
\sqrt {P_{d'} } h_{d'd} x_{d'} }  +  n_d,
\end{equation}
where $P_B$, $P_d$ and $P_{d'}$ are the transmit power of BS, D2D
transmitter $d$, $d'$, respectively. $h_{ij}$ is the channel
response of the $i-j$ link that is from equipments $i$ to $j$. $n_c$
and $n_d$ are the additive white Gaussian noise (AWGN) at the
receivers with one-sided power spectral density (PSD) $N_0$.
$\beta_{cd}$ represents the presence of interference satisfying
$\beta_{cd}=1$ when RBs of UE $c$ are assigned to UE $d$, otherwise
$\beta_{cd}=0$. As a cellular user can share resources with multiple
D2D pairs, it also satisfies $0 \le \sum\limits_d {\beta _{cd} } \le
D$. Similarly, $\beta_{dd'}$ represents the presence of interference
between D2D pairs $d$ and $d'$.

In this paper, the channel is modeled as the Rayleigh fading
channel, and thus, the channel response follows the independent
itentical complex Gaussian distribution. In addition, the free space
propagation path-loss model, $P = P_0  \cdot \left( {{d
\mathord{\left/
 {\vphantom {d {d_0 }}} \right.
 \kern-\nulldelimiterspace} {d_0 }}} \right)^{ - \alpha}$, is used where
 $P_0$ and $P$ represent signal power measured at $d_0$ and $d$ away
 from the transmitter, respectively. $\alpha$ is the path-loss exponent. Hence, the received power of
 each link can be expressed as
\begin{equation}
P_{r,ij}  = P_i  \cdot h_{ij}^2  = P_i  \cdot \left( {d_{ij} }
\right)^{ - \alpha}  \cdot h_0^2,
\end{equation}
where $P_{r,ij}$ and $d_{ij}$ are the received power and the
distance of the $i-j$ link, respectively. $P_i$ represents the
transmit power of equipment $i$, and $h_0$ is the complex Gaussian
channel coefficient that obeys the distribution $\mathcal
{C}\mathcal {N}(0,1)$. Besides, we simplify the received power at
$d_0=1$ equals the transmit power.

\subsection{System Sum Rate}
For the purpose of maximizing the network capacity, the signal to
interference plus noise ratio (SINR) should be considered as an
important indicator. The SINR of user $j$ is
\begin{equation}\label{eq:SINR}
\gamma _j  = \frac{{P_i h_{ij}^2 }}{{P_{{\mathop{\rm int}} ,j}  +
N_0 }},
\end{equation}
where $P_{{\mathop{\rm int}} ,j}$ denotes the interference signal
power received by user $j$, and $N_0$ accounts for the terminal
noise at the receiver.

Determined by the Shannon capacity formula, we can calculate the
channel rate corresponding to the SINR of cellular and D2D users. As
cellular users suffer interference from D2D communicating that
sharing the same spectrum resource, the interference power of
cellular user $c$ is
\begin{equation}\label{eq:P_int,c}
P_{{\mathop{\rm int}} ,c}  = \sum\limits_d {\beta _{cd} P_d
h_{dc}^2}.
\end{equation}
While the interference of D2D receiver $d$ is from both BS and D2D
users that are assigned the same resources to, the interference
power of user $d$ can be expressed as
\begin{equation}\label{eq:P_int,d}
P_{{\mathop{\rm int}} ,d}  = P_B h_{Bd}^2  + \sum\limits_{d'} {\beta
_{dd'} P_{d'} h_{d'd}^2 }.
\end{equation}

Based on (\ref{eq:SINR}), (\ref{eq:P_int,c}), and
(\ref{eq:P_int,d}), we can obtain the channel rate of cellular user
$c$ and D2D receiver $d$ as
\begin{equation}\label{eq:R_c}
R_c  = \log _2 \left( {1 + \frac{{P_B h_{Bc}^2 }}{{\sum\limits_d
{\beta _{cd} P_d h_{dc}^2 }  + N_0 }}} \right),
\end{equation}
\begin{equation}\label{eq:R_d}
R_d  =  \log _2 \left( {1 + \frac{{P_d h_{dd}^2 }}{{P_B h_{Bd}^2 +
\sum\limits_{d'} {\beta _{dd'} P_{d'} h_{d'd}^2 }  + N_0 }}}
\right),
\end{equation}
respectively. Here, $d \ne d'$. So $\sum\limits_{d'} {{\beta
_{dd'}}{P_{d'}}{h_{d'd}^2}}$ represents the interference from the
other D2D pairs that share spectrum resources with pair $d$.

The DL system sum rate can be defined as
\begin{equation}\label{eq:Re}
\Re  = \sum\limits_{c = 1}^C {\left( {R_c  + \sum\limits_{d = 1}^D
{\beta _{cd} R_d } } \right)}.
\end{equation}
In the next section, we formulate the problem of designing $\beta_{cd}$ for each D2D pair
as an optimization issue of maximizing $\Re$.

\section{Problem Formulation}\label{sec:prob_form}
In this section, we introduce two concepts: valuation model and
utility function, which are bases of the auction mechanism. Also,
some definitions are given.
\subsection{Valuation Model}
As D2D communication shares the same spectrum resources with
cellular communication at the same time slot, the co-channel
interference should be limited as much as possible to optimize the
system performance. The radio signals experience different degrees
of fading, and thus, the amount of interference depends on transmit
power and spatial distances. Accordingly, we focus on assigning
appropriate resource blocks (RBs) occupied by cellular users to D2D
pairs in order to minimize interference to achieve a higher system
sum rate. Next, we formulate the relation between the allocation
result and the rate of the shared channel. The relation can be
defined as a value function whose target value is the channel rate.

We define $\mathcal{D}$ as a package of variables representing the
index of D2D pairs that share the same resources. We assume the
total pairs can form $N$ such packages. Thus, if the members of the
$k$-th ($k=1,2,...,N$) D2D user package share resources with
cellular user $c$, the channel rates of UE $c$ and D2D pair $d$ ($d
\in \mathcal{D}_k$) can be written as
\begin{equation}\label{eq:R_ck}
{R_c^k} = {\log _2}\left( {1 +
\frac{{{P_B}h_{Bc}^2}}{{\sum\limits_{d \in \mathcal{D}_k}
{{P_d}h_{dc}^2}  + {N_0}}}} \right),
\end{equation}

\begin{equation}\label{eq:R_dk}
{R_d^k}  =  {\log _2} \left( {1 +
\frac{{{P_d}h_{dd}^2}}{{{P_B}h_{Bd}^2 + \sum\limits_{d' \in
\mathcal{D}_k - \left\{ d \right\}} {{P_{d'}}h_{d'd}^2}  + {N_0}}}}
\right),
\end{equation}
respectively. The rate of the operating channel shared by UE $c$ and
D2D pairs $d \in \mathcal{D}_k$ is
\begin{equation}\label{eq:R}
{R_{ck}} = {R_c^k} + \sum\limits_{d \in \mathcal{D}_k} {{R_d^k}}.
\end{equation}

According to (\ref{eq:R_ck}) $\sim$ (\ref{eq:R}), when assigning
resources of UE $c$ to the $k$-th package of D2D pairs, the channel
rate is given by
\begin{equation}\label{eq:V_ck}
V_c(k)=\log _2 \left( {1 + \frac{{P_B h_{Bc}^2 }}{{\sum\limits_{d
\in \mathcal {D}_k } {P_d h_{dc}^2 }  + N_0 }}} \right) +
\sum\limits_{d \in \mathcal {D}_k } {\log _2 \left( {1 + \frac{{P_d
h_{dd}^2 }}{{P_B h_{Bd}^2  + \sum\limits_{d' \in \mathcal {D}_k  -
\left\{ d \right\}} {P_{d'} h_{d'd}^2 }  + N_0 }}} \right)}.
\end{equation}

In the proposed reverse I-CA mechanism, we consider spectrum
resources occupied by cellular user $c$ as one of the bidders who
submit bids to compete for the packages of D2D pairs, in order to
maximize the channel rate. It is obvious that there would be a gain
of channel rate owing to D2D communicating as long as the
contribution to data signals from D2D is larger than that to
interference signals. Considering the constraint of a positive
value, we define the performance gain as
\begin{equation}\label{eq:value}
v_{c}(k) =max\left(V_{c}(k) - V_{c},0\right),
\end{equation}
which is the private valuation of bidder $c$ for the package of D2D
pairs $\mathcal{D}_k$. Here, $V_{c}$ denotes the channel rate of UE
$c$ without co-channel interference and is obtained by
\begin{equation}\label{eq:V_c}
V_c= {\log _2}\left( {1 + \frac{{{P_B}h_{Bc}^2}}{{{N_0}}}} \right).
\end{equation}
Thus, we have the following definition:

\textbf{Definition 1}: A \textbf{\emph{valuation~model}} $\mathcal
{V} = \left\{ {v_c \left( k \right)} \right\}$ is a set of the
private valuations of all bidders $c \in \left\{ {1,2, \ldots ,C}
\right\}$ for all packages $\mathcal{D}_k \subseteq \left\{ {1,2,
\ldots ,D} \right\}$ ($k \in \left\{ 1,2, \ldots ,N \right\}$).

\subsection{Utility Function}\label{subsec:utility}
In the auction, the cellular resource denoted by $c$ obtains a gain
by getting a package of D2D communications. However, there exists
some cost such as control signals transmission and information feedback
during the access process. We define the cost as a pay price.

\textbf{Definition 2}: The price to be payed by the bidder $c$ for
the package $\mathcal{D}_k$ is called \textbf{\emph{pay~price}}
denoted by $\mathcal {P}_c(k)$. The unit price of item $d$ ($\forall
k$, $d \in \mathcal{D}_k$) can be denoted by $p_c(d)$.

Here, we consider linear anonymous prices \cite{Ref:22}, which means
the prices are linear if the price of a package is equal to the sum
of the prices of its items, and the prices are anonymous if the
prices of the same package for different bidders are equal. Thus, we
have
\begin{equation}\label{eq:price}
\mathcal {P}_c(k)= \sum\limits_{d \in \mathcal {D}_k} {p_c \left( d
\right)}  = \sum\limits_{d \in \mathcal {D}_k} {p\left( d \right)}
,\forall c = 1,2, \ldots ,C.
\end{equation}
Therefore, the payment of a bidder is determined by the unit price
$p(d)$ and the size of bidding package $\mathcal {D}_k$.

\textbf{Definition 3}: \textbf{\emph{Bidder~utility}}, or named
\textbf{\emph{bidder~payoff}} $\mathcal {U}_c(k)$ expresses
satisfaction of bidder $c$ getting package $\mathcal {D}_k$. The
bidder utility can be defined as
\begin{equation}\label{eq:U}
\mathcal {U}_c(k)= v_c \left( k \right) -\mathcal {P}_c(k).
\end{equation}

Based on (\ref{eq:value}), (\ref{eq:price}), (\ref{eq:U}), $V_c(k)$
in (\ref{eq:V_ck}) and $V_c$ in (\ref{eq:V_c}), we can obtain the
utility of bidder $c$ as
\begin{align}\label{eq:utility}
\mathcal {U}_c (k) = &\log _2 \left( {1 + \frac{{P_B h_{Bc}^2
}}{{\sum\limits_{d \in \mathcal {D}_k } {P_d h_{dc}^2 }  + N_0 }}}
\right) \hspace{-0.1cm}+\hspace{-0.1cm} \sum\limits_{d \in \mathcal
{D}_k } {\log _2 \left( {1 + \frac{{P_d h_{dd}^2 }}{{P_B h_{Bd}^2  +
\sum\limits_{d' \in \mathcal {D}_k  - \left\{ d \right\}} {P_{d'}
h_{d'd}^2 }  + N_0 }}} \right)}
\nonumber\\
&- \log _2 \left( {1 + \frac{{P_B h_{Bc}^2 }}{{N_0 }}} \right) -
\sum\limits_{d \in \mathcal {D}_k } {p\left( d \right)}.
\end{align}

In order to describe the allocation outcome intuitively, we give the
definition below.

\textbf{Definition 4}: The result of the auction is a spectrum
allocation denoted by $\mathcal {X}=\left(X_1 ,X_2 , \ldots ,X_C
\right)$, which allocates a corresponding package to each bidder.
And the allocated packages may not intersect ($\forall i,j,~X_i \cap
X_j = \emptyset$).

We consider a set of binary variables $\left\{ x_c(k)\right\}$ to
redefine the allocation as
\begin{equation}\label{eq:x_ck}
x_c \left( k \right) = \left\{ {\begin{array}{*{20}c}
   {1}, & {\mbox{if}~X_c  = \mathcal {D}_k},  \\
   {0}, & {\mbox{otherwise}}.  \\
\end{array}} \right.
\end{equation}

According to the literature, two most popular bidding languages are
exclusive-OR (XOR), which allows a bidder to submit multiple bids
but at most one of the bids can win, and additive-OR (OR), which
allows one to submit multiple bids and any non-intersecting
combination of the bids can win. We consider the XOR bidding
language in this paper. Thus, (\ref{eq:x_ck}) satisfies $
\sum\nolimits_{k = 1}^N {x_c \left( k \right)}  \le 1$ and
$\sum\nolimits_{k = 1}^N {x_c \left( k \right)}  = 0 \Rightarrow
X_c= \emptyset$ for $\forall c=1,2, \ldots ,C$. If given an
allocation $\mathcal {X}$, the total bidder utility of all bidders
can be denoted as $\mathcal {U}_{all}(\mathcal {X})=\sum\nolimits_{c
= 1}^C {\sum\nolimits_{k = 1}^N {x_c(k)\mathcal {U}_c(k)}}$.
Furthermore, the auctioneer revenue is denoted as $\mathcal
{A}(\mathcal {X})=\\ \sum\nolimits_{c = 1}^C {\sum\nolimits_{k =
1}^N {x_c(k)\mathcal {P}_c(k)}} $, which is usually considered to be
the auctioneer's gain.

%

\section{Resource Allocation Algorithm Based on Reverse Iterative Combinatorial Auction}\label{sec:algorithm}
In this section, we formulate the resource allocation for D2D
communication as a reverse I-CA game. First, we introduce some
concepts of the I-CA games. Then, we investigate details of the
allocation process.

\subsection{Reverse Iterative Combinatorial Auction Game}
As mentioned before, we assume the total spectrum resources are
divided into $C$ units with each one already providing communication
service to one cellular user. By the auction game, the spectrum
units are assigned to $N$ user packages $\left\{
\mathcal{D}_1,\mathcal{D}_2,...,\mathcal{D}_N \right\}$, with each
package consisting of at least one D2D pair. In other words, the
spectrum units compete to obtain D2D communication for improving the
channel rate.

During an I-CA game, the auctioneer announces an initial price for
each item, and then, the bidders submit to the auctioneer their bids
at the current price. As long as the demand exceeds the supply, or
on the contrary that the supply exceeds the demand, the auctioneer
updates (raises or reduces) the corresponding price and the auction
goes to the next round.

Obviously, it can be shown that the overall gain, which includes the
total gain of the auctioneer and all bidders does not depend on the
pay price, but equals to the sum of the allocated packages'
valuations, i.e.,
\begin{align}
\mathcal {A}\left( \mathcal {X} \right) + \mathcal {U}_{all} \left(
\mathcal {X} \right) &= \sum\limits_{c = 1}^C {\sum\limits_{k = 1}^N
{x_c \left( k \right)\mathcal {P}_c \left( k \right)} } +
\sum\limits_{c = 1}^C {\sum\limits_{k = 1}^N {x_c
\left( k \right)\mathcal {U}_c \left( k \right)} } \nonumber\\
&= \sum\limits_{c = 1}^C {\sum\limits_{k = 1}^N {x_c \left( k
\right)\mathcal {P}_c \left( k \right)} }  + \sum\limits_{c = 1}^C
{\sum\limits_{k = 1}^N {x_c \left( k \right)[\left( {v_c \left( k
\right) - \mathcal {P}_c \left( k \right)} \right)]} } \nonumber\\
&= \sum\limits_{c = 1}^C {\sum\limits_{k = 1}^N {x_c \left( k
\right)v_c \left( k \right)} }.
\end{align}

As our original intention, we employ the I-CA to obtain an efficient
allocation for spectrum resources.

\textbf{Definition 5}: An \textbf{\emph{efficient allocation}}
denoted by $\tilde{\mathcal
{X}}=(\tilde{X}_1,\tilde{X}_2,\ldots,\tilde{X}_C)=\left\{
\tilde{x}_c(k) \right\}$ is an allocation that maximizes the overall
gain.

Given the private bidder valuations for all possible packages in
(\ref{eq:value}), an efficient allocation can be obtained by solving
the combinatorial allocation problem (CAP).

\textbf{Definition 6}: The \textbf{\emph{Combinatorial Allocation
Problem (CAP)}}, also sometimes referred as \textbf{\emph{Winner
Determination Problem (WDP)}}, leads to an efficient allocation by
maximizing the overall gain: $\mathop {\max }\limits_{\mathcal {D}_k
= X_c  \in \mathcal {X} \in \mathscr{X}} \sum\nolimits_{c = 1}^C
{v_c \left( k \right)}$, where $\mathscr{X}$ denotes the set of all
possible allocations.

An integer linear program using the binary decision variables
$\left\{ x_c(k)\right\}$ is formulated for the CAP as
\begin{align}\label{eq:CAP}
&\max \sum\limits_{c = 1}^C {\sum\limits_{k = 1}^N {x_c \left( k
\right)v_c \left( k \right)} }, \\
s.t.~~~&\sum\limits_{k=1}^N {x_c(k)} \leq 1, \forall c \in \left\{
1,2, \ldots, C \right\}, \nonumber\\
&\sum\limits_{\mathcal {D}_k :d \in \mathcal {D}_k } {\sum\limits_{c
= 1}^C {x_c (k)}  \le 1} , \forall d \in \left\{ 1,2, \ldots, D
\right\}. \nonumber
\end{align}
The objective function maximizes the overall gain, and the
constraints guarantee: 1) at most one package can be allocated to
each bidder; 2) each item cannot be sold more than once.

In fact, there might be multiple optimal solutions of the CAP with
the same objective function. From the auctioneer's point of view,
tie-breaking rules are needed to determine which of the optimal
solutions is selected. In a real auction, the auctioneer does not
know the private valuations of the bidders, neither can it solve the
NP-hard problem. To solve the CAP, the auctioneer selects the
winners on the basis of the submitted bids in each round. Therefore,
in case of the XOR bidding language, the WDP formulation is similar
to the CAP and the only difference is the objective function
\begin{equation}\label{eq:WDP}
\max \sum\limits_{c = 1}^C {\sum\limits_{k = 1}^N {x_c \left( k
\right)\mathcal {P}_c^t \left( k \right)} },
\end{equation}
where $\mathcal {P}_c^t(k)$ represents the pay price of bidder $c$
for package $\mathcal {D}_k$ in round $t$.

Based on Definition 5, the overcome of a CA is not always efficient.
Here, we employ allocating efficiency as a primary measure to
benchmark auctions.

\textbf{Definition 7}: \textbf{\emph{Allocating efficiency}} in CAs
can be expressed as the ratio of the overall gain of the final
allocation to that of an efficient allocation \cite{Ref:22}
\begin{equation}
\mathcal {E}(\mathcal X)=\frac{{\mathcal {A}\left( \mathcal{X}
\right) + \mathcal{U}_{all} \left( \mathcal{X} \right)}}{{\mathcal
{A} \left(\tilde{\mathcal {X}} \right) + \mathcal{U}_{all} \left(
\tilde {\mathcal {X}} \right)}},
\end{equation}
which has $\mathcal{E}(\mathcal X) \in [0,1]$.

\subsection{Algorithm for Resource Allocation}\label{subsec:algorithm}
In this subsection, the details of the allocation scheme based on
reverse I-CA are introduced. We has modeled the D2D resource
allocation problem as a reverse I-CA game and gave the valuation
model, utility function and other important concepts. Many I-CA
designs, especially for the centralized I-CA design, are based on
ask prices. The price-based I-CA designs differ by the pricing
scheme and price update rules. In the proposed algorithm, linear
prices are used as mentioned in Subsection \ref{subsec:utility} for
they are easy to understand for bidders and convenient to
communicate in each auction round. Because of the interference from
D2D links, cellular channels should guarantee the performance of
cellular system before allowing the D2D access. Hence, we consider a
descending price criterion in the algorithm. Prices update by a
greedy mode that once a bidder submits a bid for items or packages
the corresponding prices are fixed, otherwise the prices are
decreased.

At the beginning of the allocation, the BS collects the location
information of all the D2D pairs. In addition, the round index
$t=0$, the initial ask price $p^0(d)$ for each item (D2D pair) $d$,
and the fixed price reduction $\Delta>0$ are set up. When the
initial prices are announced to all the bidders (i.e. spectrum
resources occupied by cellular UEs), each bidder submits bids, which
consist of its desired packages and the corresponding pay prices.
Jump bidding where bidders are allowed to bid higher than the
prices, is not allowed in our scheme, thus bidders always bid at the
current prices. According to the CAP proposed in Definition 6 and
the analysis about the WDP, we simplify the problem of maximizing
the overall gain as a process of collecting the highest pay price.
As a result, bidder $c$ bids for package $\mathcal {D}_k$ as long as
$\mathcal{U}_c(k)\geq 0$. Combining (\ref{eq:price}) and
(\ref{eq:U}), we have
\begin{equation}\label{eq:bid_condition}
v_c \left( k \right) \ge \mathcal{P}_c^t \left( k \right) =
\sum\limits_{d \in \mathcal{D}_k } {p^t \left( d \right)},
\end{equation}
where the round index $t \ge 0$. In this case, let $b_c^t(k)=\left\{
\mathcal{D}_k,\mathcal{P}_c^t(k) \right\}$ denote the submitted bid
at the end of round $t$, and $\mathcal{B}^t=\left\{
b_c^t(k)\right\}$ denotes all the bids. When
(\ref{eq:bid_condition}) is not satisfied, bid $b_c^t(k)=\left\{
\emptyset, 0 \right\}$.

If $\exists d \in \mathcal{D}_k$ satisfies $\forall b_c^t(k) \in
\mathcal{B}^t, \mathcal{D}_k \notin b_c^t(k)$, it reveals that the
supply exceeds the demand. Then, the BS sets $t=t+1$,
$p^{t+1}(d)=p^t(d)-\Delta$ where $d$ is the over-supplied item, and
the auction moves on to the next round.

In a normal case, as long as the price of a package decreases below
a bidder's valuation for that package, the bidder submits a bid for
it. The BS allocates the package to the bidder, and fixes the
corresponding prices of items. At the same time, constrained by the
XOR bidding language, the bidder is not allowed to participate the
following auction rounds. As the asking prices decrease discretely
every round, it may exist a situation that more than one bidders bid
for packages containing the same items simultaneously. The BS
detects the bids of all the bidders: 1) it exists
$b_{c_1}^t(k)=b_{c_2}^t(k)\neq\left\{ \emptyset,0\right\}$ ($c_1
\neq c_2,k\in \left\{1,2,\ldots,N \right\}$); 2) it exists
$b_{c_1}^t(k_1)=\left\{
\mathcal{D}_{k_1},\mathcal{P}_{c_1}^t(k_1)\right\}$,
$b_{c_2}^t(k_2)=\left\{
\mathcal{D}_{k_2},\mathcal{P}_{c_2}^t(k_2)\right\}$ ($k_1\neq
k_2,c_1,c_2\in \left\{ 1,2,\ldots,C\right\}$) satisfying
$\mathcal{D}_{k_1}\cap\mathcal{D}_{k_2}\neq \emptyset$. If either of
the above conditions is satisfied, the overall demand exceeds supply
for at least one item. Then, the BS sets a fine tuning
$p^t(d)=p^t(d)+\delta$ where $d$ is the temporary over-demanded
item, and $\delta$ can be set by $\delta=\Delta/i$ where $i$ is an
integer factor that affects the convergence rate. The allocation can
be determined by multiple iterations.

The auction continues until all the D2D links are auctioned off or
every channel wins a package. Our algorithm is detailed in
Table~\ref{table_1}.

\section{Analysis of the Proposed Resource Allocation Algorithm}\label{sec:analysis}
In this section, we investigate the important properties of the
proposed auction-based resource allocation mechanism.
\subsection{Cheat-Proof}
As the general definition, cheat-proof means that reporting the true
demand in each auction round is a best response for each bidder.

\textbf{Proposition 1}: The resource allocation algorithm based on
the reverse I-CA is cheat-proof.

\begin{proof} From (\ref{eq:utility}), we can get that the utility
of bidder $\mathcal{U}_c(k)$ depends on the valuation of the package
it bids and unit prices of the items. In details, it is the
interference (between cellular and D2D communications) that mainly
affects the utility. As the expression is extremely complex to
resolve, we consider the case that only one item constitutes the
package without loss of generality. The utility of bidder $c$ can be
rewritten as
\begin{equation}
\mathcal{U}_c (d) = \log _2 \left( {1 + \frac{{P_B h_{Bc}^2 }}{{P_d
h_{dc}^2 + N_0 }}} \right) + \log _2 \left( {1 + \frac{{P_d h_{dd}^2
}}{{P_B h_{Bd}^2  + N_0 }}} \right) - \log _2 \left( {1 + \frac{{P_B
h_{Bc}^2 }}{{N_0 }}} \right) - p^t\left( d \right),
\end{equation}
and the differential expressions of the utility with respect to
$h_{dc}$ and $h_{Bd}$ are
\begin{equation}\label{eq:differential_1}
\frac{{\partial \mathcal{U}_c \left( d \right)}}{{\partial h_{dc} }}
= \frac{{ - 2P_d h_{dc} P_B h_{Bc}^2 }}{{\ln 2\left( {P_d h_{dc}^2 +
P_B h_{Bc}^2 + N_0 } \right)\left( {P_d h_{dc}^2  + N_0 } \right)}}
< 0,
\end{equation}

\begin{equation}\label{eq:differential_2}
\frac{{\partial \mathcal{U}_c \left( d \right)}}{{\partial h_{Bd} }}
= \frac{{ - 2P_B h_{Bd} P_d h_{dd}^2 }}{{\ln 2\left( {P_B h_{Bd}^2 +
P_d h_{dd}^2 + N_0 } \right)\left( {P_B h_{Bd}^2  + N_0 } \right)}}
< 0,
\end{equation}
respectively. Accordingly, utility $\mathcal{U}_c(d)$ is a
monotonically decreasing function with respect to both $h_{dc}$ and
$h_{Bd}$. Thus, the optimal strategy is to bid the D2D link that has
a lower channel gain with the cellular transmitter and receiver.

In a descending price auction, items are always too expensive to
afford at the beginning. With the number of iterations $t$
increasing, the prices of items drop off. Given a package
$\mathcal{D}_k$ in round $t$, bidder $c$ has the right to submit bid
$\left\{ \mathcal{D}_k,\mathcal{P}_c^t(k)\right\}$ or $\left\{
\emptyset,0 \right\}$. Given that all the other bidders submit their
true demands according to (\ref{eq:bid_condition}), we consider the
strategy of bidder $c$ in two cases: 1) if $c$ bids $\left\{
\emptyset,0 \right\}$ when its true valuation for $\mathcal{D}_k$
satisfies $\mathcal{U}_c(k) \geq 0$, it will quit this round and
lose the package which maximizes its channel rate; 2) if $c$ bids
$\left\{ \mathcal{D}_k,\mathcal{P}_c^t(k)\right\}$ when its true
valuation for $\mathcal{D}_k$ satisfies $\mathcal{U}_c(k) < 0$ and
finally wins the package, it will obviously get a negative surplus
that is unwanted.

From the above analysis, we can conclude that the optimal strategy
for cellular channel $c$ is to submit its true demand in each round,
or it will get a loss in its utility as a result of any deceiving.
That is, the proposed resource allocation algorithm is cheat-proof.
\end{proof}

\subsection{Convergence}\label{subsec:conv}
In this subsection, we prove that the proposed algorithm has the
convergence property.

\textbf{Proposition 2}: The resource allocation algorithm based on
the reverse I-CA has the convergence property that the number of the
iterations is finite.

\begin{proof}: According to Theorem 1, all the bidders submit their
true demands in each auction round, in order to obtain the utility
from winning. From (\ref{eq:utility}), we can derive
\begin{equation}
\mathcal{U}_c^{t+1}-\mathcal{U}_c^{t}=\Delta>0,
\end{equation}
where $\mathcal{U}_c^{t}$ denotes the utility of bidder $c$ in round
$t$. According to the algorithm, we have that bidder $c$ will get
zero utility with no bid if $\mathcal{U}_c^t<0$, and have an
opportunity to win a positive utility with bid $\left\{
\mathcal{D}_k,\mathcal{P}_c^t(k)\right\}$ if $\mathcal{U}_c^t \geq
0$. Therefore, in the beginning, bidder $c$ plays a waiting game,
and once $\mathcal{U}_c^t(k) \geq 0$, it will bid for
$\mathcal{D}_k$. As long as it is the only one that submits a bid,
it will get the package. With a sufficiently large $t$ and
$\Delta>0$, we can finally get $x_c(k)=1$. Similarly, if more than
one bidders bid for the same item, we can have an allocation by
ascending price process with the step $\delta<\Delta$. Subjected to
$\sum\limits_{\mathcal {D}_k :d \in \mathcal {D}_k } {\sum\limits_{c
= 1}^C {x_c (k)}  \le 1}$ in (\ref{eq:CAP}), the package can not be
sold once more. Thus, for a finite number of packages $N$, the
number of iterations is finite. That is, the proposed scheme would
reach convergence.
\end{proof}

In addition, the value of the price step $\Delta$ has a direct
impact on the speed of convergence of the proposed scheme. The
scheme converges fast when $\Delta$ is large, while it converges
slowly when $\Delta$ is small. The fine tuning $\delta$ also has the
same nature, but less impact on convergence.

\subsection{Price Monotonicity}
In an I-CA game, the price updates through several ways, i.e.,
monotonically increasing, monotonically decreasing and non-monotonic
modes. Here, we focus on the price non-monotonicity in the proposed
reverse I-CA algorithm.

\textbf{Proposition 3}: In the proposed descending price auction,
the raising item prices in a round may be necessary to reflect the
competitive situation. Moreover, it brings efficiency improvement.

\begin{proof} From the algorithm proposed in Table \ref{table_1}, there exists a situation that more than one
bidders submit bids for the same package or different packages with
intersection when prices are reduced to some certain values. But
auctions do not allow one item being obtained by multiple bidders as
the second constraint in (\ref{eq:CAP}) shows. In this situation,
raising the corresponding prices by a fine tuning $\delta=\Delta/i$
makes bidders to reinspect their utility functions. Once a bidder
finds its utility less than zero, it quits from the competition. By
a finite number of iterations, the winner converges to one bidder.
Since the ascending price process maximizes the auctioneer revenue
as shown in (\ref{eq:WDP}), the allocation has higher efficiency
than a random allocation in that situation.
\end{proof}


\subsection{Complexity}
As mentioned before, a traditional CAP in fact is an NP-hard
problem, the normal solution of which is the centralized exhaustive
search. We set that the number of items to be allocated is $m$, and
the number of bidders is $n$. For an exhaustive optimal algorithm,
an item can be allocated with $n$ possible results. Thus, all the
$m$ items are allocated with $n^m$ possible results. The complexity
of the algorithm can be denoted by $\mathcal {O}(n^m)$. In the
proposed reverse I-CA scheme, bidders reveal their entire utility
function, i.e., they calculate valuations for all possible packages,
the number of which is
$\mathcal{C}_m^1+\mathcal{C}_m^2+\cdots+\mathcal{C}_m^m=2^m - 1$. If
the total number of iterations is $t$, the complexity of the
auction-based scheme is $\mathcal{O}(n\left( {2^m - 1} \right)+t)$.
From the proposed algorithm, we have $p^t \left( d \right) = p^0
\left( d \right) - \Delta  \cdot t \ge 0 $ (The fine tuning has a
small impact on the result and can be omitted here). So the worst
case is $t = {{p^0 \left( d \right)} \mathord{\left/
 {\vphantom {{p^0 \left( d \right)} \Delta }} \right.
 \kern-\nulldelimiterspace} \Delta }$. It is obvious that for sufficient large values of $m$ and $n$,
general values of $p^0(d)$ and $\Delta$, a much lower complexity is
obtained by using the proposed reverse I-CA scheme. That is,
$\mathcal{O}(n^m)>\mathcal{O}(n\left(2^m-1\right)+{{p^0 \left( d
\right)} \mathord{\left/
 {\vphantom {{p^0 \left( d \right)} \Delta }} \right.
 \kern-\nulldelimiterspace} \Delta })$. If we constrain the
number of D2D pairs sharing the same channel to one, the complexity
would be further reduced to $\mathcal{O}(n \cdot m+{{p^0 \left( d
\right)} \mathord{\left/
 {\vphantom {{p^0 \left( d \right)} \Delta }} \right.
 \kern-\nulldelimiterspace} \Delta })$. And the performance of this
reduced scheme is included in the simulation in Subsection VI-A.

\subsection{Overhead}
In D2D underlay system, the BS is still the control center of
resource allocation, and the global CSI should indeed be available
at the BS for the proposed scheme. In addition to the CSI detection,
feedback, and the control signaling transmission, the reverse I-CA
scheme does not need additional signaling overhead compared to
existing resource scheduling schemes such as maximum carrier to
interference (Max C/I) and proportional fair (PF), which also need
the global CSI to optimize the system performance. The difference is
that the reverse I-CA scheme requires more complicated CSI due to
the interference between D2D and cellular network.


At the beginning of the allocation, the transmitters need to send
some packets containing detection signals. Then, the obtained CSI at
each terminal (D2D or cellular receiver) would be feedback to the
BS. After that, iteration process would be conducted at the BS, and
no signaling needs to be exchanged among the network nodes until the
control signals forwarding.

Methods, such as CSI feedback compression and signal flooding, would
help reduce the overhead. In addition, the future work on D2D
communication could consider some mechanism that limit the number of
D2D pairs sharing the same channel by, e.g. distance constraint,
which would obviously help reduce the overhead. But for this paper,
the target is to obtain the nearest-optimal solution, wherefore we
do not consider the simplification.

\section{Simulation Results and Discussions}\label{sec:simulations}
In this section, we provide the simulation results to illustrate the
performances of the proposed reverse I-CA algorithm. Besides, we
give the necessary analysis for the results. The main simulation
parameters are listed in Table \ref{table_2}. As shown in Fig.
\ref{fig:1}, simulations are carried out in a single cell. Both
path-loss model and shadow fading are considered for cellular and
D2D links. The wireless propagation is modeled according to WINNER
II channel models \cite{Ref:WINNER}, and D2D channel is based on
office/indoor scenario while cellular channel is based on the urban
microcell scenario.

\subsection{System Sum Rate}
\begin{figure}[h!]
\centering
\includegraphics[height=3.7in]{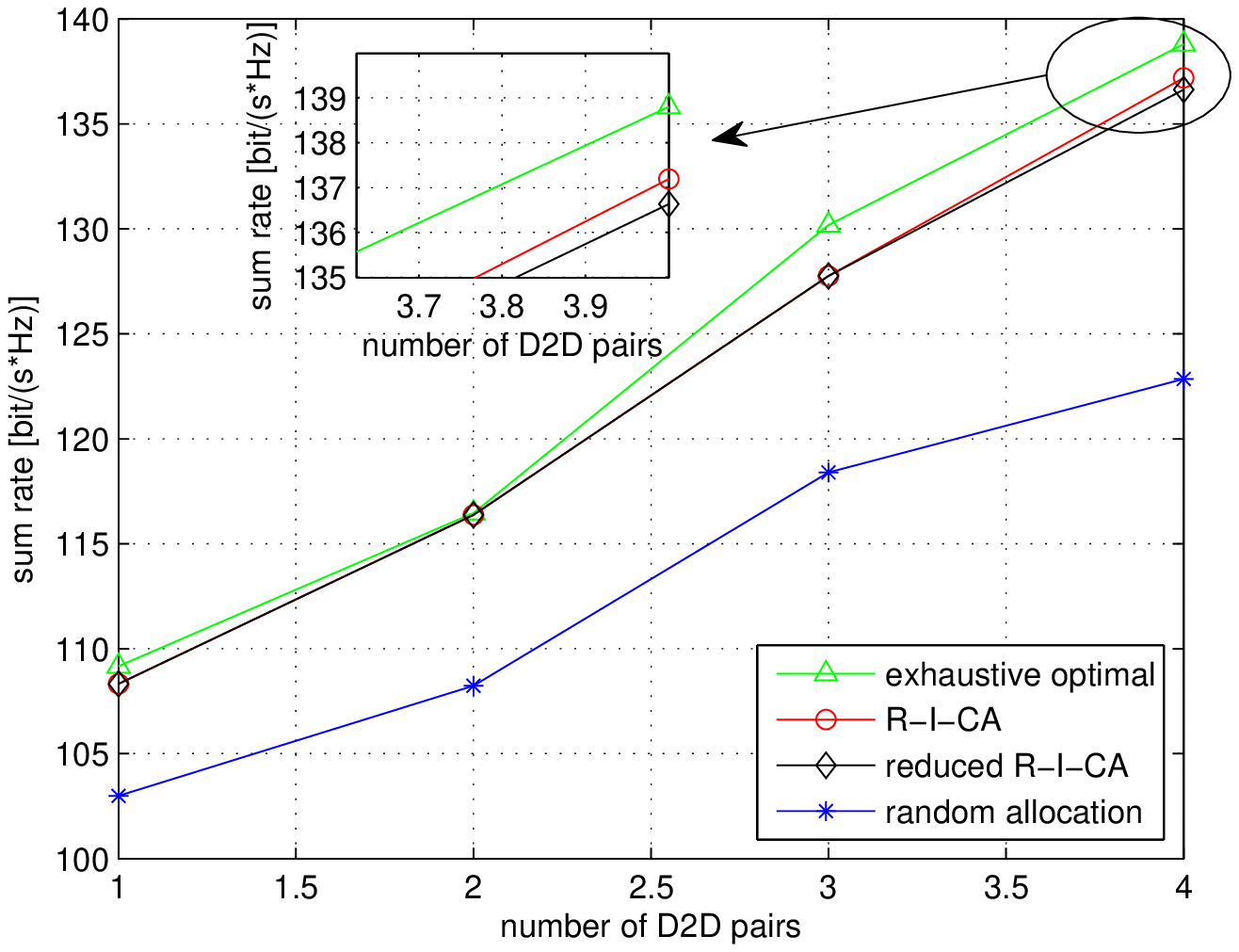}
\caption{System sum rate for different allocation algorithms in the
case of 8 resource units.} \label{fig:2}
\end{figure}

\begin{figure}[h!]
\centering
\includegraphics[height=3.7in]{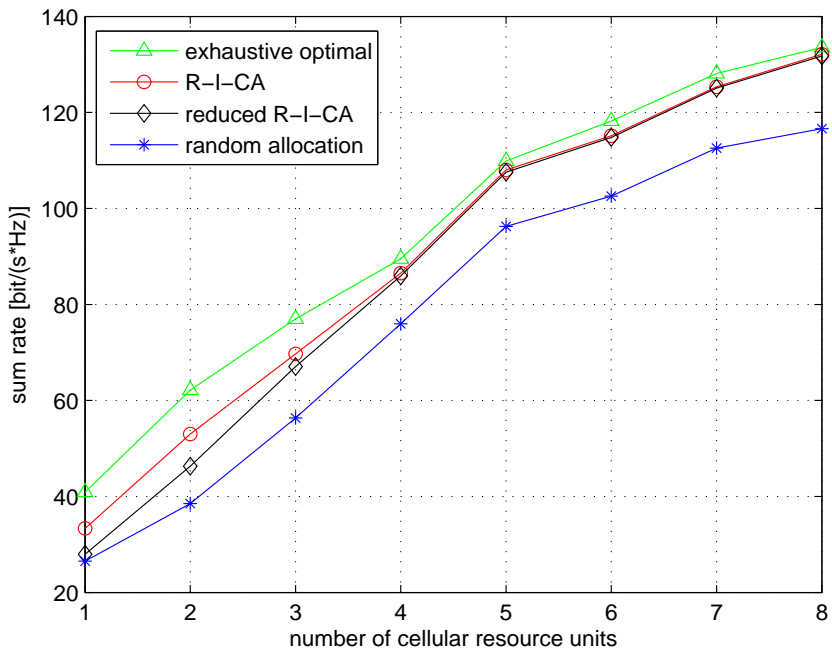}
\caption{System sum rate for different allocation algorithms in the
case of 4 D2D pairs.} \label{fig:4}
\end{figure}

The system sum rate with different numbers of D2D pairs and
different numbers of resource units using the proposed auction
algorithm is illustrated in Fig. \ref{fig:2} $\sim$ Fig.
\ref{fig:4}. The sum rate can be obtained from (\ref{eq:Re}).

From Fig. \ref{fig:2} and Fig. \ref{fig:4}, we can see that the
system sum rate goes up with both the number of D2D pairs and the
number of resource units increasing. On one side, when the amount of
resources is fixed, more D2D users contribute to a higher system sum
rate. On the other side, as the amount of resource increases, the
probability of resources with less interference to D2D links being
assigned to them enhances, which can lead to the increased sum rate.
This phenomenon is similar to the effect of multiuser diversity.
Definitely, cellular users also contributes to the performance.

\begin{figure}[h!]
\centering
\includegraphics[height=3.7in]{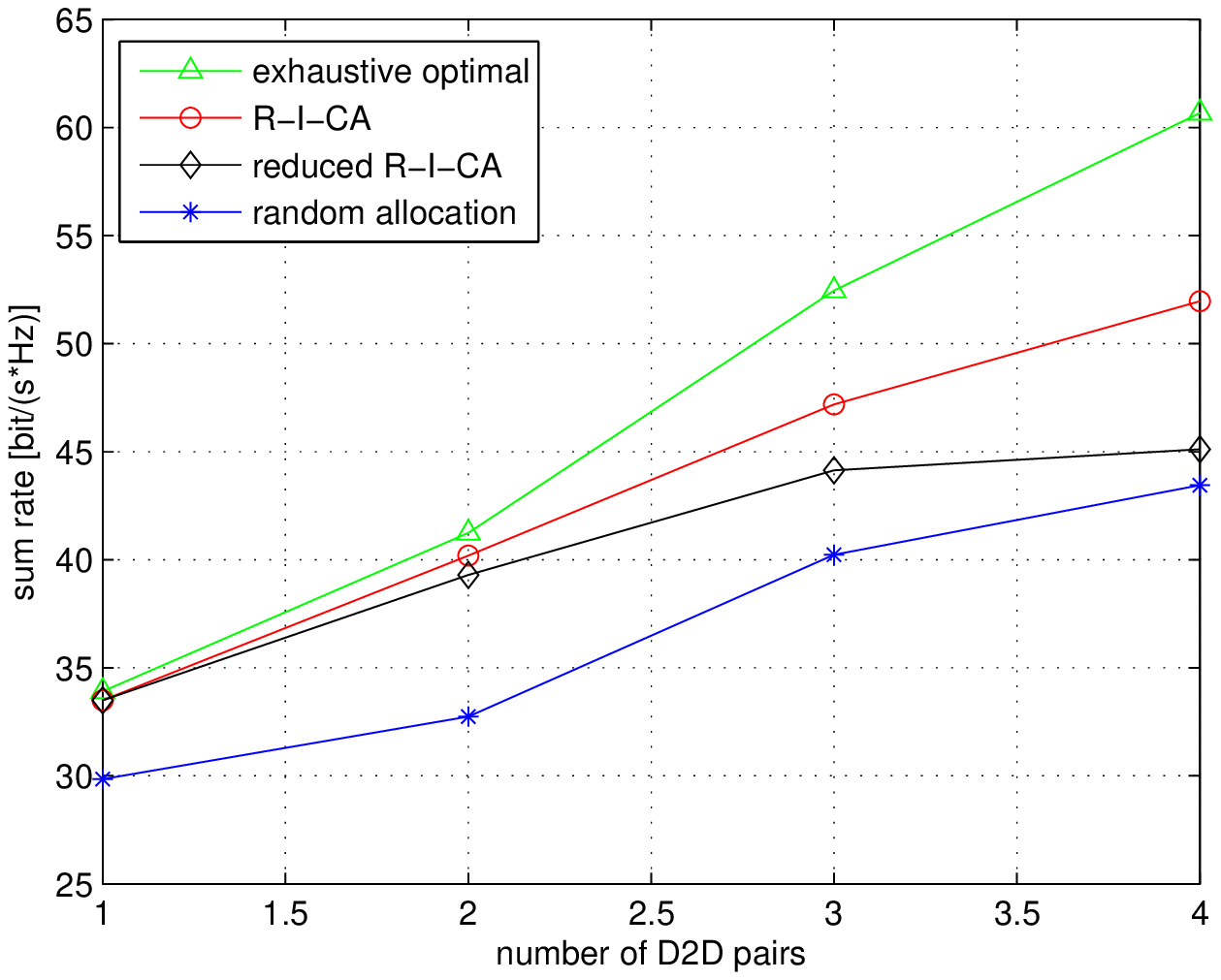}
\caption{System sum rate for different allocation algorithms in the
case of 2 resource units.} \label{fig:3}
\end{figure}

From another perspective, Fig.\ref{fig:2} $\sim$ Fig. \ref{fig:4}
shows the system sum rate for different allocation algorithms. The
curve marked exhaustive optimal is simulated by the exhaustive
search way, which guarantees a top bound of the system sum rate. The
curve marked reduced R-I-CA is the result of a reduced reverse I-CA
scheme, in which the number of D2D pairs sharing the same cellular
resources is constrained to one. The curve marked R-I-CA represents
the performance of the proposed reverse I-CA algorithm, and the last
one is the simulation result using random allocation of spectrum
resources. Firstly, we can see that the proposed auction algorithm
is relatively much superior to the random allocation. Secondly, the
optimal allocation results in the highest system sum rate, but the
superiority compared to R-I-CA is quite small, especially when the
number of cellular resource units increases as Fig. \ref{fig:4}
shows. Moreover, we find that the performance of reduced R-I-CA
approximates to that of R-I-CA scheme in case of 8 resource units,
but differs obviously in case of 2 resource units shown in Fig.
\ref{fig:3}. The reason for this phenomenon is that the constraint
of the reduced R-I-CA limits D2D pairs accessing to the network when
the number of resources units is less than that of D2D pairs, thus a
large capacity loss products.

\subsection{System Efficiency}
\begin{figure}[h!]
\centering
\includegraphics[height=3.7in]{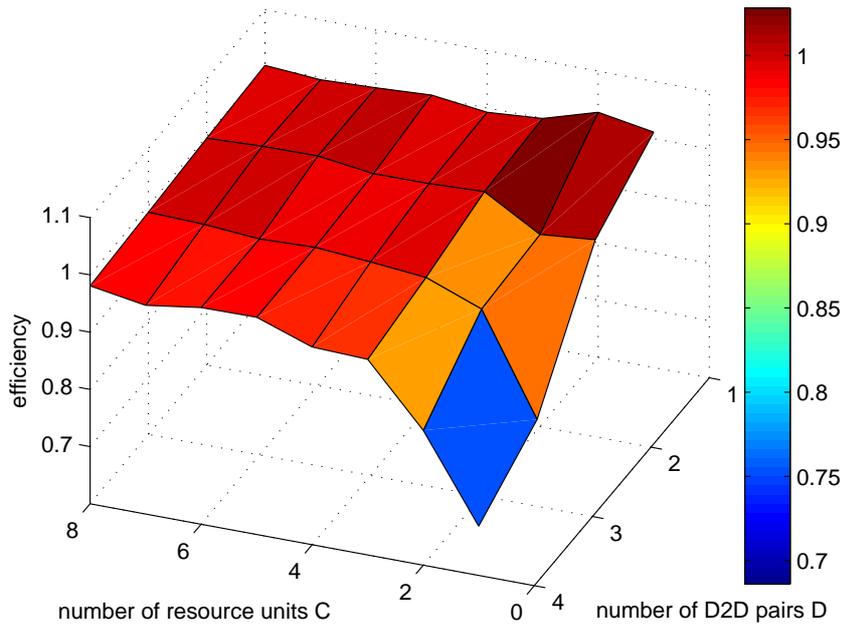}
\caption{System efficiency: $\eta$ with different numbers of D2D
pairs and different numbers of resource units.} \label{fig:5}
\end{figure}

We define the system efficiency as $\eta  = {\Re  \mathord{\left/
 {\vphantom {\Re  {{\Re _{opt}}}}} \right.
 \kern-\nulldelimiterspace} {{\Re _{opt}}}}$, where $\Re_{opt}$
 represents the exhaustive optimal sum rate.
Fig. \ref{fig:5} shows the system efficiency with different numbers
of D2D pairs and different numbers of resource units. The simulation
result indicates that the proposed algorithm provides high (the
lowest value of $\eta$ is around 0.7) system efficiency. Moreover,
the efficiency is stable over different parameters of users and
resources.

As to the point of efficiency value being about 0.7, the number of
resource units and the number of D2D pairs are both small. The
linear price rule limits bidders to bid the maximal valuation
packages, but to bid the packages having maximal average unit
valuation. For this reason, the efficiency decreases slightly.

As to other points, the efficiency is stable above 0.9, which
reflects a small performance gap between the proposed algorithm and
the exhaustive search scheme. In fact, the descending price rule
determines the bidder that has the highest bid on current items
would win the corresponding package, which maximizes the current
overall gain. However, the gap cannot be avoided as the algorithm
essentially follows a local, or an approximate global optimum
principle.

\subsection{Price Monotonicity}\label{subsec:monotonicity}
\begin{figure}[h!]
\centering
\includegraphics[height=3.7in]{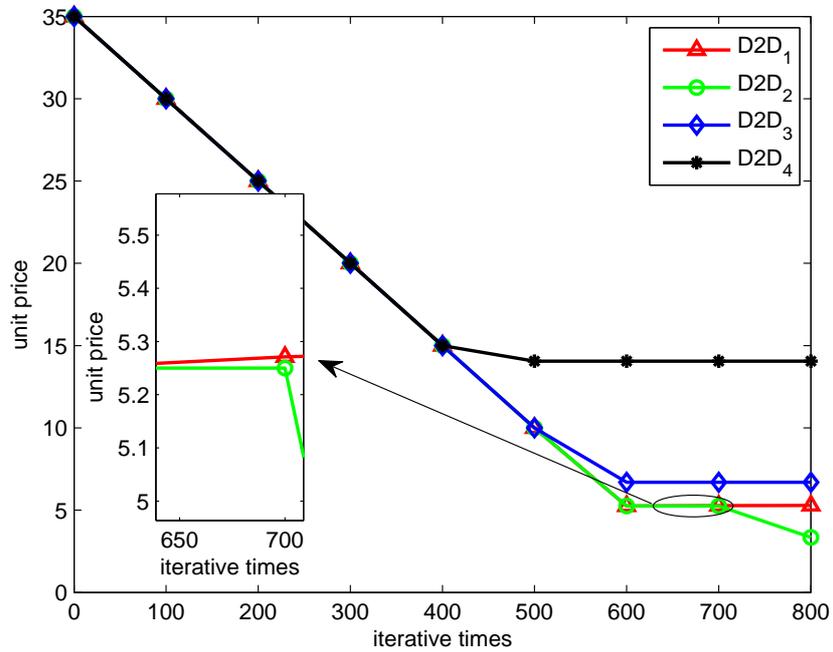}
\caption{Price monotonicity: an example of price non-monotonicity in
the reverse I-CA scheme.} \label{fig:6}
\end{figure}

Fig. \ref{fig:6} shows an example of the price non-monotonicity in
the reverse I-CA scheme. The four curves represent unit price of
four D2D pairs. As the enlarged detail shows, the unit price of D2D
pair $2$ has an ascending process during the auction. As the step
$\delta$ is much less than descending step $\Delta$, the phenomenon
of ascending price is hard to pick out. When the items have been
sold out, their prices are fixed to the selling value. And from the
figure, we can find that the D2D pair $2$ is the last one to be
sold.


\section{Conclusions}\label{sec:conclusions}
In this paper, we have investigated how to reduce the effects of
interference between D2D and cellular users, in order to improve the
system sum rate for a D2D underlay network. We have proposed the
reverse iterative combinatorial auction as the mechanism to allocate
the spectrum resources for D2D communications with multiple user
pairs. We have formulated the valuation of each D2D pair for each
resource unit, and then explained a detailed auction algorithm
depending on the utility function. A non-monotonic descending price
iteration process has been modeled and analyzed to be cheat-proof,
converge in a finite number of rounds, and has low complexity. The
simulation results show that the system sum rate goes up with both
the number of D2D pairs and the number of resource units increasing.
The proposed auction algorithm is much superior to the random
allocation, and provides high system efficiency, which is stable
over different parameters of users and resources.

\renewcommand{\baselinestretch}{1.4}

\newpage
\begin{table}[!t]
\renewcommand{\arraystretch}{1.3}
\caption{The resource allocation algorithm} \label{table_1}
\centering
\begin{tabular}{p{130mm}}
\hline

$\ast$ \textbf{Initial State:}

\quad The BS collects the location information of all D2D pairs. The
valuation of the $c$-th resource unit for package $k$ is $v_c(k), c
= 1,2,\ldots,C, k = 1,2,\ldots,N$, which is given by
(\ref{eq:value}). The round index
$t=0$, and the initial price $P^0(d)$, the fixed price reduction $\Delta >0$ are set up.\\

$\ast$ \textbf{Resource Allocation Algorithm:}

1. Bidder $c$ submits bids $\left\{
\mathcal{D}_k,\mathcal{P}_c(k)\right\}$ depending on its utility.

\quad $\star$ bidder $c$ bids for package $\mathcal{D}_k$ as long as
$\mathcal{U}_c(k)\geq0$, which is represented by
(\ref{eq:bid_condition}).

\quad $\star$ If $\mathcal{U}_c(k)<0$, bidder $c$ submits $\left\{
\emptyset,0 \right\}$.

2. If $\exists d \in \mathcal{D}_k$ satisfies $\forall b_c^t(k) \in
\mathcal{B}^t, \mathcal{D}_k \notin b_c^t(k)$, the BS sets $t=t+1$,
$p^{t+1}(d)=p^t(d)-\Delta$ where $d$ is the over-supplied item, and
the auction moves on to the next round. Return to step 1.

3. The BS detects the bids of all the bidders:

\quad \quad 1) it exists $b_{c_1}^t(k)=b_{c_2}^t(k)\neq\left\{
\emptyset,0\right\}$ ($c_1 \neq c_2,k\in \left\{1,2,\ldots,N
\right\}$);

\quad \quad 2) it exists $b_{c_1}^t(k_1)=\left\{
\mathcal{D}_{k_1},\mathcal{P}_{c_1}^t(k_1)\right\}$,
$b_{c_2}^t(k_2)=\left\{
\mathcal{D}_{k_2},\mathcal{P}_{c_2}^t(k_2)\right\}$ ($k_1\neq
k_2,c_1,c_2\in \left\{ 1,2,\ldots,C\right\}$) satisfying
$\mathcal{D}_{k_1}\cap\mathcal{D}_{k_2}\neq \emptyset$.

4. If neither of the conditions in step 3 is satisfied, go to step
5. Otherwise, the overall demand exceeds supply for at least one
item. The BS sets $p^t(d)=p^t(d)+\delta$, and $\delta$ can be set by
$\delta=\Delta/i$ where $i$ is an integer factor. Return to step 1.

5. The allocation can be determined by repeating the above steps.
The auction continues until all D2D links are auctioned off
or every cellular channel wins a package.\\

\hline
\end{tabular}
\end{table}

\begin{table}
\begin{center}
\caption{Main Simulation Parameters}\label{table_2}
\begin{tabular}{|l|l|}
\hline \bf{Parameter} & \bf{Value}\\ \hline Cellular layout &
Isolated cell, 1-sector\\ \hline System area & The radius of the
cell is 500 m\\ \hline  Noise spectral density & -174 dBm/Hz\\
\hline Sub-carrier bandwidth & 15 kHz \\ \hline Noise figure & 9 dB at device\\
\hline Antenna gains & BS: 14 dBi; Device: 0 dBi\\ \hline
The maximum distance of D2D & 5 m\\
\hline Transmit power & BS: 46 dBm; Device: 23 dBm \\ \hline
\end{tabular}
\end{center}
\end{table}

\end{document}